\documentclass[useAMS,usenatbib,usegraphicx]{mn2e}
\usepackage{amsmath}
\usepackage{amssymb}
\usepackage{soul}
\title [Hungarias as possible source of co-orbital bodies.]{Hungaria region as possible source of Trojans and satellites in the
inner solar-system.}
\author[M. A. Galiazzo and R. Schwarz]
{M. A. Galiazzo$^{1}$\thanks{E-mail:mattia.galiazzo@univie.ac.at}, R. Schwarz$^{1}$\\
$^{1}$Institute of Astrophysics, University of Vienna, A-1180 Vienna, 
T\"urkenschanzstrasse 17, Austria\\ 
}

\begin{document}

\date{Accepted 2014 September 15. Received 2014 July 7; in original 
form 2014 October 11}

\pagerange{\pageref{firstpage}--\pageref{lastpage}} \pubyear{2002}
\maketitle
\label{firstpage}

\begin{abstract}

The Hungaria Family (the closest region of the Main Belt to Mars)
 is an important source of Planet-Crossing-Asteroids and even
 impactors of terrestrial planets. We present the possibility that asteroids
 coming from the Hungaria Family get captured into co-orbital motion with the
 terrestrial planets in the inner solar system.
Therefore we carried out long term numerical integrations (up to
 100 Myr) to analyze the migrations from
 their original location - the Hungaria family region- into the
 inner solar system.
During the integration time we observed whether or not the Hungarias
 get captured into a co-orbital motion with
by the terrestrial planets.
Our results show that 5.5 \% of 200 Hungarias,
 selected as a sample of the whole
 group, escape from the Hungaria region and
 the probability from that to become
 co-orbital objects (Trojans, satellites or horseshoes)
 turns out to be $\sim$ 3.3\%:  1.8\% for Mars and 1.5\% for the Earth.
 In addition we distinguished in which classes of co-orbital motion
the asteroids get captured and for how long they stay there in stable
 motion. Most of the escaped Hungarias become Quasi-satellites
 and the ones captured as Trojans favour the $L_5$ lagrangian
point.
This work highlights that the Hungaria region is a source
 of Mars and also Earth co-orbital objects.

\end{abstract}


\section{Introduction}
\label{intro}

A co-orbital configuration refers to a celestial object (such as an asteroid)
that keeps a quasi-constant distance from its parent object (in this work,
 planet) and it is on a $1:1$ mean motion resonance (MMR).
 In this configuration
 the asteroid has a rotational period around the Sun similar to the planet
 which is co-orbiting.
 
The co-orbital bodies are subdivided in classes of objects
 which depend on their point of libration. 
In this study we are interested in the following classes
 for the Inner solar system:
 (a) Trojan objects, which librate around one of the two stable Lagrangian
 equilibrium points,
 $L_4$ and $L_5$, respectively asteroids leading (libration angle
 $\lambda \sim +60^\circ$) and heading ($\lambda \sim -60^\circ$) the
 planets orbit,
 i.e. 2010 TK$_7$ for the Earth ~\citep{Con2011} and
 (b) Satellites (MOs) and quasi-satellites (QSs) orbits,
 which librates around 0$^\circ$, but the libration width $\sigma$ is 
much larger for the QSs (more details are presented in 
Section~\ref{binsetup}).
In contrast to MOs, QSs orbits 
lie outside the planet's Hill sphere, therefore they are not long-term stable.
Over time they tend to evolve to other types of resonant motion, where they no 
longer remain in the planet's neighborhood.

Currently one Earth Trojan and 9 Martian Trojan asteroids, 5 horseshoe
 objects (one Martian and 4 of the Earth)
 and also 6 quasi satellite close to the Earth, are known. All known 
co-orbital objects -- including candidates -- in the inner solar
system are presented in Tab.~\ref{tabREAL}.
Theoretical studies predict that Trojan asteroids are a byproduct of 
planet formation and evolution and were later captured from the planets.
Chaotic capture of Jovian Trojan asteroids in the early Solar System
 ($\sim$3.4 My), were presented in the work of \cite{Mor2013}.
~\cite{Lyk2009} and ~\cite{Lyk2010} investigated the origin and dynamical
 evolution of Neptune Trojans during the formation and migration
 of the planets. 
They found that the captured Trojans display a wide range of inclinations 
($0^{\circ}\lesssim i < 40^{\circ}$). 
These results were confirmed by~\cite{Sch2012}, who investigated the
capture probability of co-orbital objects for the planets Venus, Earth
and Mars.

Early work on the origin of NEAs \citep[e.g.][]{Gre1989,Gre1993},
 suggested that collisions in the main-belt 
continuously produce new asteroids by fragmentation of larger bodies. 
These fragments can be injected into the $\nu_6$ and J3:1 MMR
 with Jupiter, which causes a change of their 
eccentricities and brings them into orbits 
intersecting the orbits of Mars (Mars crossers) and/or Earth
 ~\cite[Earth crossers, e.g.][]{Mil1989}:
 gravitationally, the NEAs are transported first to Mars, mainly by MMRs,
three-body mean motion resonances ~\cite[3BMMRs, for a description of this
 kind of resonances see][]{Nes1998}
 and secular resonances (SRs), and then to other more interior planets 
due especially to close encounters with Mars.
 Also non-gravitational forces can play a role in the
 transportation as was shown by \cite{Bot2002,Bot2006,Gre2012}
 and \cite{Cuk2014}, but as a first stage we will take into account only
 gravitational forces in this work and the Yarkovsky effect will be
 considered in a future work.\\
In order to describe the primordial main belt before the LHB,
 a hypothetical inner extension of the main belt\footnote{The
 primordial main belt before the Late Heavy Bombardment (LHB) event},
 has been suggested and dubbed the ``E-belt'' ~\citep{Bot2012}.
One motivation for this inference is to provide a source for
 basin-forming lunar impacts of the LHB.
 These E-belt asteroids were supposed to
 have a  semi-major axis ranging from 1.7 to 2.1 au.
 Prior to the giant planet migration described in the
 Nice model, these asteroids would have been in a more
 stable orbit, with the $\nu_6$ secular resonance outside the
 border of this region ~\citep{Mor2010} with
 the outer giant planets having a  more compact configuration
 with an almost circular orbit ~\citep{Gom2005}.
 Then, during the migration of the giant planets
 ~\citep{Min2011}, the $\nu_6$ and other related
 resonances would have destabilized the E-belt population.
 Most of them would have moved inward onto
 terrestrial planets as their eccentricities and inclinations
 increased making impacts with the planets and
 so some of these asteroids (0.1-0.4 \%) would have
 acquired orbits similar to the Hungarias. In this sense,
 the Hungarias are supposed to be a remnant of the E-belt;
 the survivors of the E-belt dispersion \citep{Bot2012}. 
This idea is a development of the NICE model ~\cite[see in
 particular][]{Mor2010}  and it should make it more consistent.
The NICE model has still some gaps, the most important are:
 (a) it does not explain the presence of Mercury and (b)
 the rate of the incoming comets and
 even an explanation of the large-scale mixing of reddish
 and bluish material (from the photometric
 point of view) in the asteroid belt ~\citep{DeM2014}.
 For this reason we study in this work only the present
 Hungaria group, which might be an evolution of the ancient E-belt.\\
The importance of considering Hungarias as source of NEAs (which can originate
 also possible co-orbital bodies of terrestrial planets),
 is shown very well in \cite{Gal2013a} and in \cite{Cuk2014},
 who described the dynamical evolution of these mainly E-type 
asteroids \citep{Car2001,Ass2008,War2009} into the NEAs region.\\
Here we perform a numerical study
 on the orbits of the asteroids of the 
Hungaria Family (see also \cite{Gal2013a}),
  investigating their capture probability
  into the 1:1 MMR with the terrestrial planets: Venus, Earth and Mars.
Hungarias are relatively far out away from the orbit of the terrestrial planets,
 in fact 
the inner part starts with a semi-major axis equal to 1.78 au~\citep{Gal2013a}.
To study the capture of the Trojan asteroids into the inner
 Solar System it 
is necessary to consider the interactions (collisions and mass transport) 
between the Near-Earth-Asteroids (NEAs) and the main-belt asteroids.

\begin{table}
\centering
 \caption{All observed Earth and Mars co-orbital asteroids. *
   depicts an object which is only a candidate. The different motion types are horseshoe orbits $H$ and tad-pole 
orbits in Lagrangian points $L_4$  and $L_5$ or in both of this last two 
consecutively, like jumping Trojans $JT$. $T_j$ 
represents the Tisserand parameter in respect of Jupiter and $MT$ stands for
 motion type.}
  \begin{tabular}{llllll}
  \hline
 Name       & a [au]   & e  & i [$^\circ$] & $T_j$ & MT\\
  \hline
&{\bf Mars}&&&\\
\hline
(121514) 1999 UJ$_7$    & 1.5245 & 0.039 & 16.8 &4.449& $L_4$ \\
(5261) Eureka & 1.5235 & 0.065 & 20.3&4.428 & $L_5$ \\
(101429) 1998 VF$_{31}$   & 1.5242& 0.100 & 31.3 & 4.334& $L_5$ \\
(311999) 2007 NS$_2$    & 1.5237 & 0.054 & 18.6 &4.439 &$L_5$ \\
(269719) 1998 QH$_{56}$*  & 1.5507 & 0.031 & 32.2 &4.279 &$L_5$\\
(385250) 2001 DH$_{47}$  & 1.5238 & 0.035 & 24.4 &4.400 &$L_5$\\
2001 SC$_{191}$  & 1.5238 & 0.044 & 18.7 &4.439 &$L_5$\\
(88719) 2011 SL$_{25}$  & 1.5238 & 0.115 & 21.5 &4.415 &$L_5$\\
 2011 UN$_{63}$  & 1.5237 & 0.064 & 20.4 & 4.427 &$L_5$\\
(157204) 1998 SD$_4$    & 1.5149 & 0.125 &  13.7 &4.475 &$H$ \\
  \hline
&{\bf Earth}&&&\\
  \hline
2010 TK$_7$  & 1.0000 & 0.191 & 20.9 & 6.008& $JT$ \\
(3753) Cruithne & 0.9977 & 0.515 & 19.8 & 5.922  & $QS$\\
(164207) 2004 GU$_9$ & 1.0013 & 0.136 & 13.7&6.041 & $QS$\\ 
(277810) 2006 FV$_{35}$ & 1.0013 & 0.378 & 7.1 &6.003 &$QS$\\ 
2003 YN107 & 0.9987 & 0.014 & 4.3 &6.132 &$QS$\\ 
 (54509) YORP  & 1.0060 & 0.230 & 1.600&6.028 & $QS$\\
2001 GO$_2$ & 1.0067 & 0.168 & 4.620&6.033 & $QS$\\
2013 BS$_{45}$ & 0.9939 & 0.084 & 0.773&6.106&$H$ \\ 
2010 SO$_{16}$ & 1.0019 & 0.075 & 14.5&6.041&$H$ \\ 
2002 AA$_{29}$    & 0.9926 & 0.012 & 10.8&6.100 & $H$\\
2006 JY$_{26}$   & 1.0100 & 0.083 & 1.4&6.030 & $H$\\
(85770) 1998 UP$_1$* & 0.9983 & 0.345 & 33.2&5.901 & $H$\\
 \hline
\end{tabular}
\label{tabREAL}
\end{table}

During the integration time we observe whether or not the Hungarias
 get captured into a co-orbital motion with the planets in the inner
 Solar-system, from Venus to Mars.
 In addition we distinguish in which classes of co-orbital motion
the asteroids get captured and for how long they stay there in stable
 motion.
Therefore we carry out long term numerical integrations up to
 100 Myr to analyze the transfers from
 their original location - the Hungaria family region- towards the
 terrestrial planets.\\
The paper is organized as follows: the model and the methods are described in
Section~\ref{binsetup}; the results are shown in Section~\ref{HCOss} 
(subdivided in two subsections, subsection ~\ref{sample},
 where we describe some sample cases of Hungaria orbital evolution
 and transport mechanism and subsection ~\ref{result}, where we give 
the probability for an Hungaria to get in 
co-orbital motion with a terrestrial planet, its lifetime and orbit in such
 configuration). The conclusions are in Section~\ref{conclusions}.

\section{Model and methods}
\label{binsetup}
We do numerical N-body simulations using the Lie integration method
 ~\citep{Han1984,Egg2010,Sch2012,Gal2013a}.  
We continue the last work of~\cite{Gal2013a} considering the calculations 
of the Hungaria group: we take a sub-sample of 200 bodies, 
representative of the whole group, as the most evolved ones, selected out of 
the total sample of 8258 asteroids\footnote{The orbital data are taken from 
the ASTORB database (http://www2.lowell.edu/elgb)} considering a criterion
 based on the osculating elements. We choose the following variable 
$d=\sqrt{\Big({\frac{a}{<a>}\big)^2}\Big({\frac{e}{<e>}\big)^2}\Big({\frac{sin i}{<sin i>}\big)^2}}$ and picked up 200 Hungarias with the highest
 values of $d$.
Therefore first we integrate the orbits to the asteroids of the Hungaria group
 as defined in \cite{Gal2013a}\footnote{The Hungaria group is
 defined in this region of osculating elements: $1.78 < a [au] < 2.03$,
 $12^\circ < i < 31^\circ$ and $e<0.19$.
 A sub-sample of 200 bodies representative
 of the Hungaria group are integrated in a simplified   Solar System
 (Sun, Mars, Jupiter, Saturn and the mass-less asteroids), for 100 Myr
 to identify possible escapers, like \cite{Gal2013a}. 

 In fact analyzing the orbits of the clones of 3 Hungarias
 (100 clones per asteroid) next to resonances with the Earth and Venus
 (i.e. V1:4 and E2:5);
  including also these 2 planets in the integrations, we found only
 one important deflection out of 300 bodies.}.

Then, after the first integration, 11 fugitives out of 200 are
 detected and therefore
they are again dynamically investigated.
For any fugitives, 49 clones were generated:
 random values for (a, e, i), beginning
with the escapers' initial conditions
  in the following ranges:
 $a \pm 0.005$ (au) $e \pm 0.003$ and $i \pm 0.005^\circ$.

The model for the solar system is now from Venus to Saturn
 and the integration time is once more 100 Myr.
Finally we search for captures with the terrestrial planets
 (Venus, Earth and Mars). \\
Whenever we find a capture, we integrate again the orbit of the asteroid
 from the point when they get captured. We perform another
 integration (with the same simplified solar system) with 
a smaller\footnote{The first integration, where the Hungaria orbits 
were computed for 100 Myr, had a time step of 1000 years}
 time step (100 d) for 20 kyr and studying 
the orbit in detail.\\
The aim of this work is to study the capture of Hungaria asteroids 
in the Inner solar system, in particular for 2 different types of captures:
 1) Satellite orbits and 2) Tadpole orbits ($L_4$ and $L_5$).
 In some cases we find Horseshoes orbits  and jumping
 Trojans\footnote{Asteroids which jump from $L_4$ to $L_5$ or 
vice versa ~\citep{Tsi2000}}, too.

 The classification was done by the help of the libration width $\sigma$, 
which is defined as the difference between the mean longitude of the asteroid
and the planet (Venus, Earth or Mars) ($\lambda -\lambda_P$). $\lambda$, 
$\lambda_P$ are given by $\lambda=\varpi+M$, $\lambda_P=\varpi_P+M_P$ were 
$\varpi$, $\varpi_P$ are the longitudes of the asteroid and of the planet and 
$M$, $M_P$ are the mean anomaly of the asteroid respectively of the planet.

In a next step we compared the distributions of the orbital elements a, e, 
and i.
We also examine the orbital histories of captured objects to determine type
of capture and the orbital evolution of objects before and after a capture 
event.

\section{Hungaria co-orbital objects (HCOs)}
\label{HCOss}
There are 7 candidates (out of 11) among the Hungaria fugitives in
 \cite{Gal2013a} which can be captured
 in to co-orbital motions with terrestrial planets,
 the initial condition can be found in Table~\ref{HCOs}.

\begin{table}
\begin{center}
\begin{tabular}{lccc}
\hline
Asteroid     & $a$ [AU] & $e$ & $i$ [deg] \\
\hline
(211279) 2002 RN$_{137}$ & 1.8538 & 0.1189 & 22.82 \\
(152648) 1997 UL$_{20}$ & 1.9894 & 0.1841 & 28.88 \\
(141096) 2001 XB$_{48}$ & 1.9975 & 0.1055 & 12.32 \\
(24883) 1996 VG$_9$ & 1.8765 & 0.1556 & 22.71 \\
(41577) 2000 SV$_2$ & 1.8534 & 0.1843 & 24.97 \\
(175851) 1999 UF$_5$ & 1.9065 & 0.1874 & 19.24 \\
(39561) 1992 QA & 1.8697   & 0.1116 & 26.2 \\
(41898) 2000 WN$_{124}$* & 1.9073 & 0.1062& 17.11 \\
(30935) Davasobel* & 1.9034& 0.1178 & 27.81 \\
(171621) 2000 CR$_{58}$* & 1.9328 & 0.1051 & 17.19 \\
(129450) 1991 JM* & 1.8512& 0.1263 & 24.50 \\
\hline
\end{tabular}
\caption{Osculating elements for the escaping Hungarias: semi-major axis ($a$), eccentricity ($e$),
 inclination ($i$) in degrees; data taken from the database ``astorb.dat''. These 7 asteroids also
 belong to the Hungaria family (see astdys website, http://hamilton.dm.unipi/astdys/, for comparisons 
with the elements), so we can treat the fugitives restrictively also as member
 of the Hungaria \emph{family}. * means they are not candidate HCOs.}
\label{HCOs}
\end{center}
\end{table}
 
Among all the Hungaria fugitives we found co-orbital objects
 (from now on HCOs),
  like tadpole orbits ($L_4$ and $L_5$), MOs, QSs and
 some horseshoe orbits, too.

\subsection{Sample cases of a Trojan and of a Quasi-satellite}
\label{sample}

\begin{figure}
\centerline
{\includegraphics[width=7.0cm,angle=0]{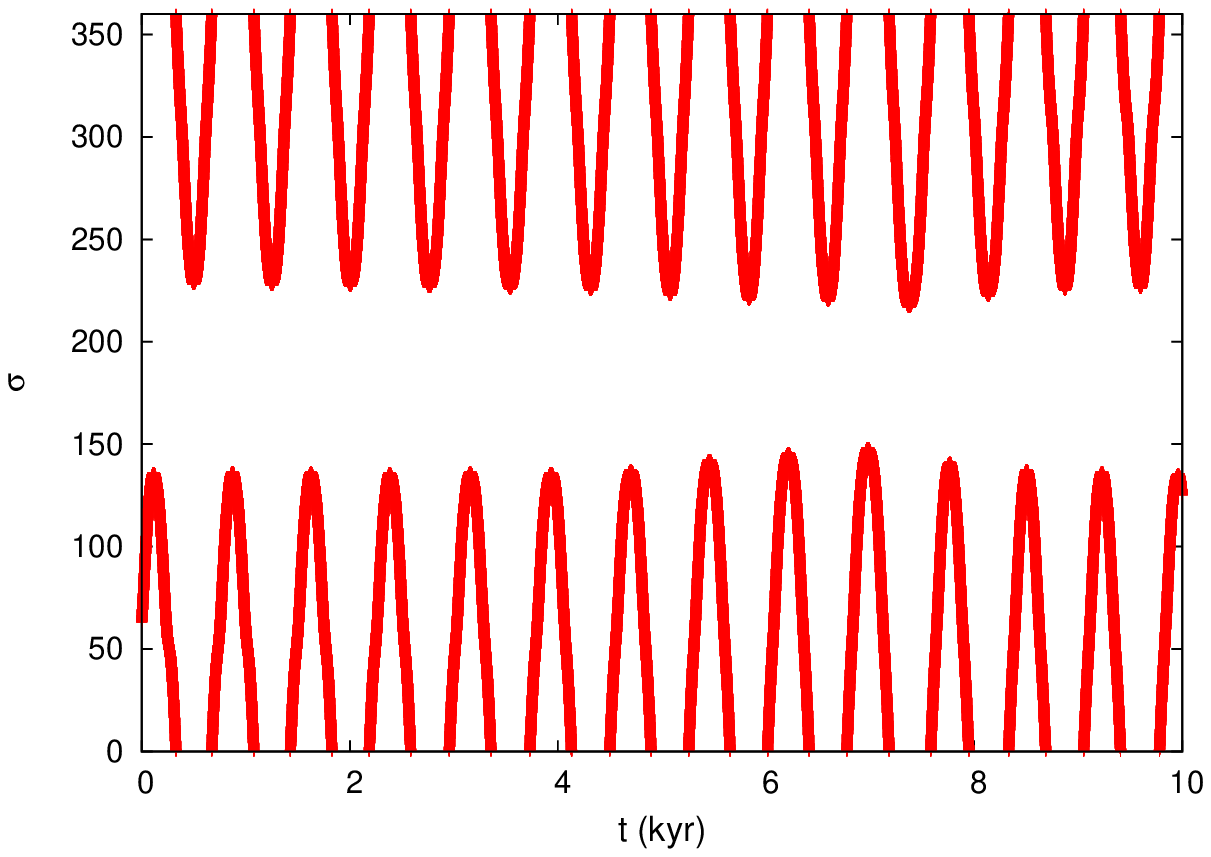}}
{\includegraphics[width=7.0cm,angle=0]{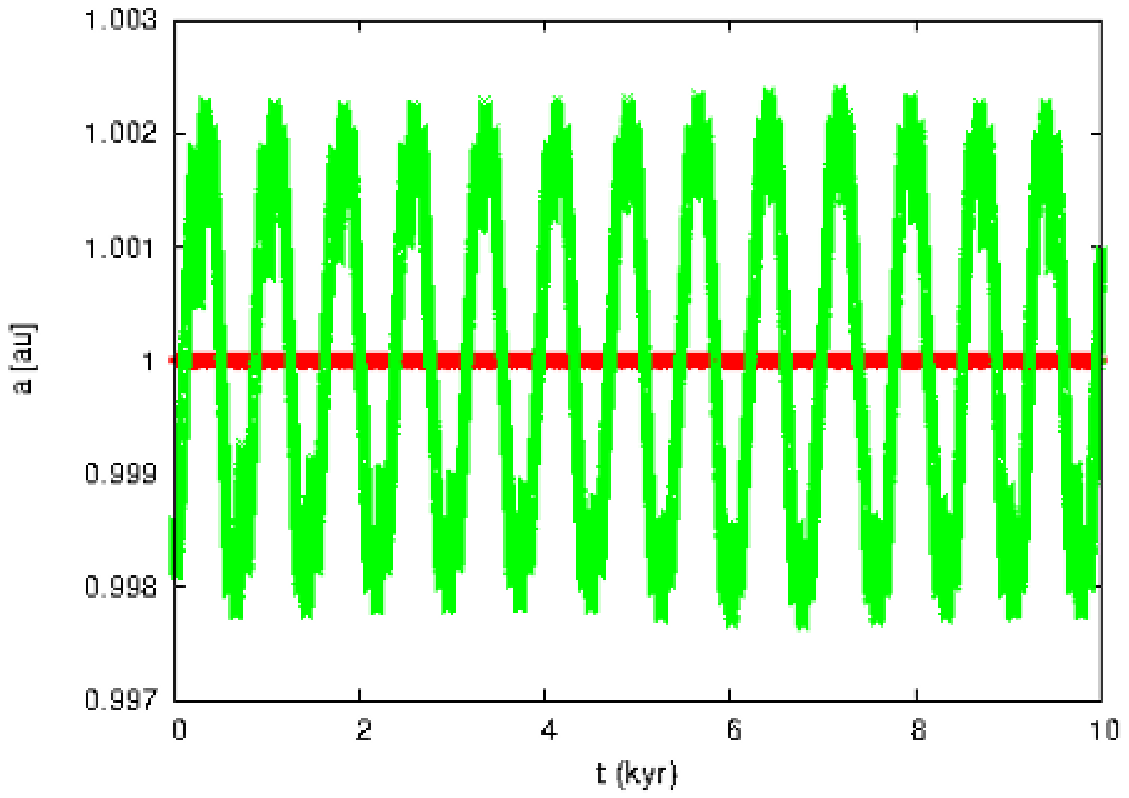}}
{\includegraphics[width=7.0cm,angle=0]{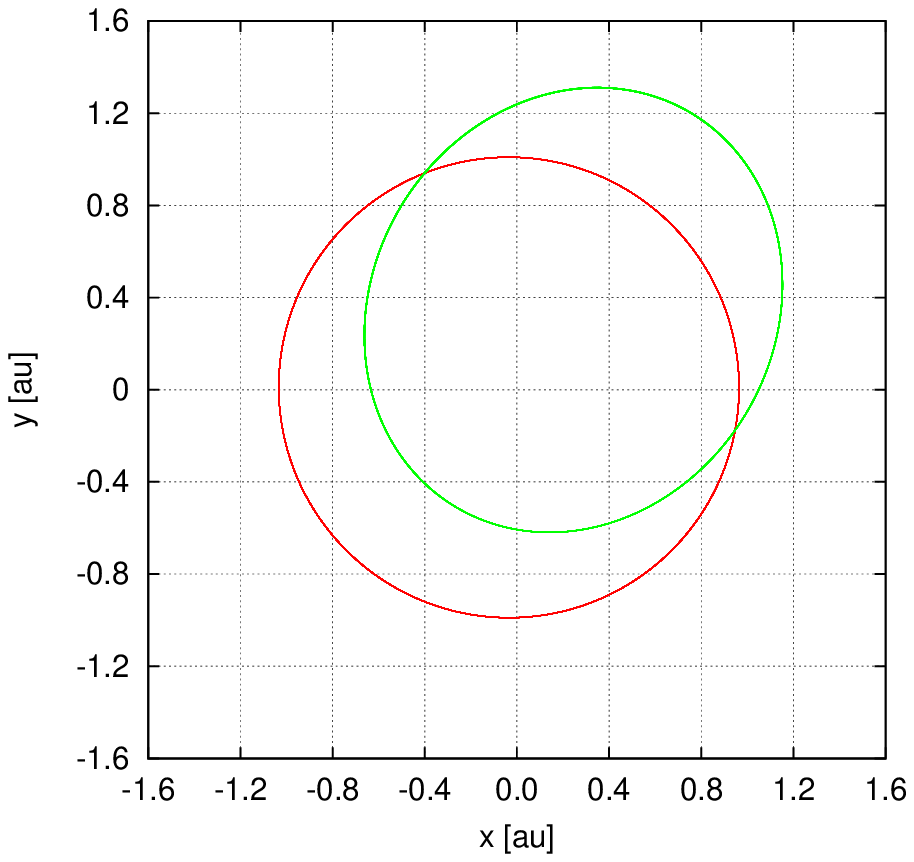}}
{\includegraphics[width=7.0cm,angle=0]{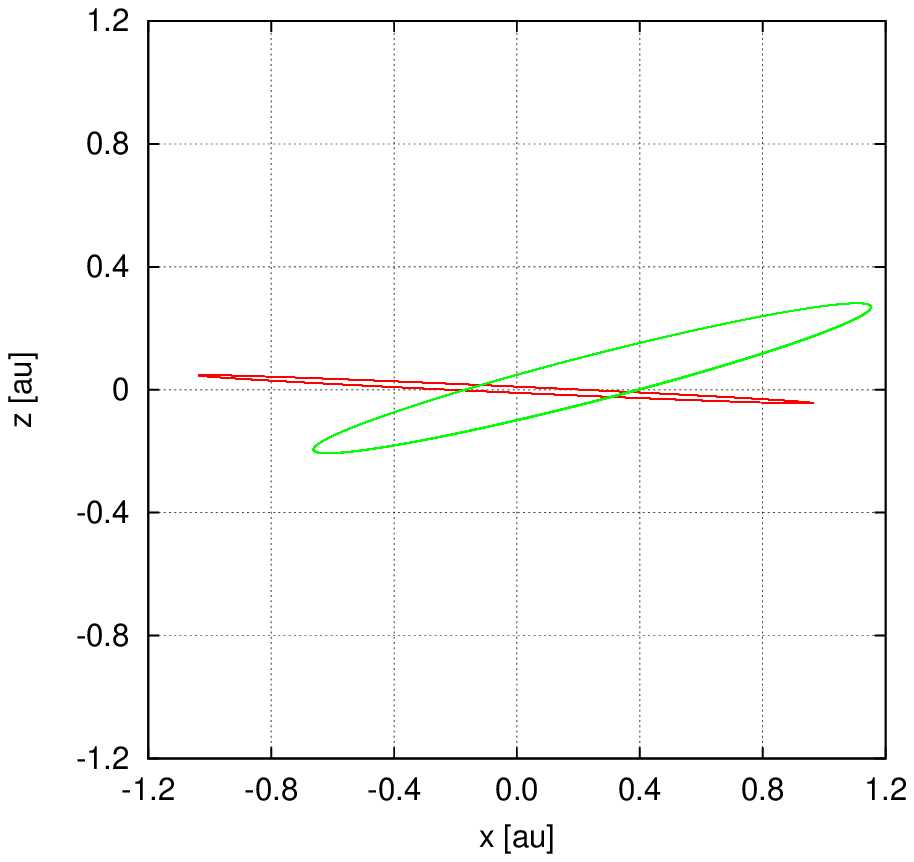}}
\caption{From top to bottom: 1) Libration angle ($\sigma$) of the Earth QS  (141096) 2001 XB$_{48}$; 2) the semi-major axis of the asteroid 
librating around the one of the Earth; 3) views of the orbits of Earth (curve with a radius of 1 au) and a clone of the asteroid 2001 XB$_{48}$ (captured by the Earth for about 10 kyr) as seen from above the north ecliptic pole in the geocentric plane, emphasizing the eccentric orbit of this quasi satellite; 4) Views of the orbits of the bodies of point (3) seen on the plane perpendicular to the ecliptic. The inclined orbit of the asteroid to the Earth is clear, allowing excursions of roughly 0.2 and 0.3 au above and below the plane.}
\label{fig12}
\end{figure}

\begin{figure}
  \centering
\includegraphics[width=7.0cm]{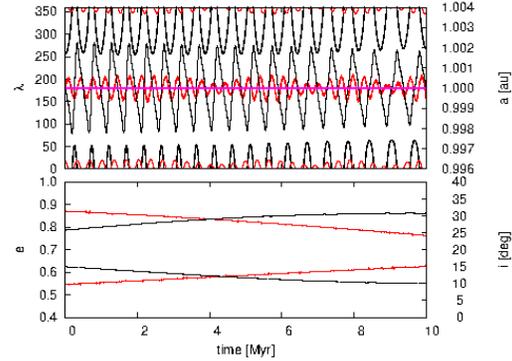}
\caption{Comparison between a terrestrial satellite and a Quasi-satellite. In the upper panel the critical angle (upper and bottom curves) versus time is represent and the semi-major axis variation (central curves, apart the horizontal line which represents the semi-major axis of the Earth) versus time again. Lower panel represents eccentricity and inclination versus time of the 2 different types of co-orbital bodies. The configurations represent the satellite-state of 2001 XB$_{48}$ (lighter color) and quasi-satellite state (darker color) in co-orbital motion with the Earth.}  
\label{moonquasimoon}
\end{figure}

We analyse the orbital evolution of the fugitive clones,
 observing whether 
they get captured in to co-orbital motion with terrestrial planets. 
We have to mention that we never find a case where an asteroid get captured
by Mars and then by the Earth together and there is no case for Venus
 co-orbital motion.
We find several cases of QSs, i.e. for a clone
 of (141096) 2001 XB$_{48}$\footnote{several clones of 
different asteroids goes in co-orbital bodies and several ones of
2001 XB$_{48}$ become QSs} 
(see the graphics description in Fig.~\ref{fig12} and ~\ref{moonquasimoon}),
 but also some jumping Trojans.

\subsubsection{Orbital evolution of a typical HCO and transport mechanism}

\begin{figure}
\centerline
{\includegraphics[width=7.0cm,angle=0]{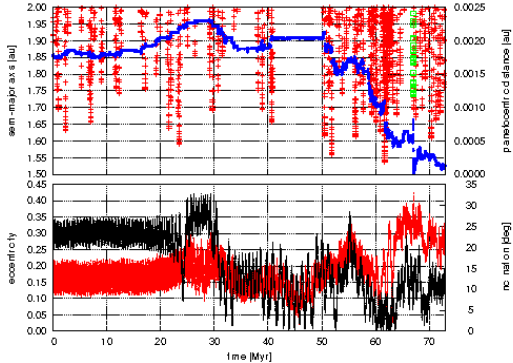}}
\caption{Evolution of the orbit till the initial instant of co-orbital motion. Upper panel: semi-major axis and Planetocentric distance versus time for a clone of the asteroid 2002 RN$_{137}$.
 In vertical points in crosses and dot-quadrate, close
encounters with respectively: Mars and the Earth. Bottom panel: eccentricity (light color) and inclination (black).}
\label{2002aei}
\end{figure}

\begin{figure}
\centerline
{\includegraphics[width=7.0cm,angle=0]{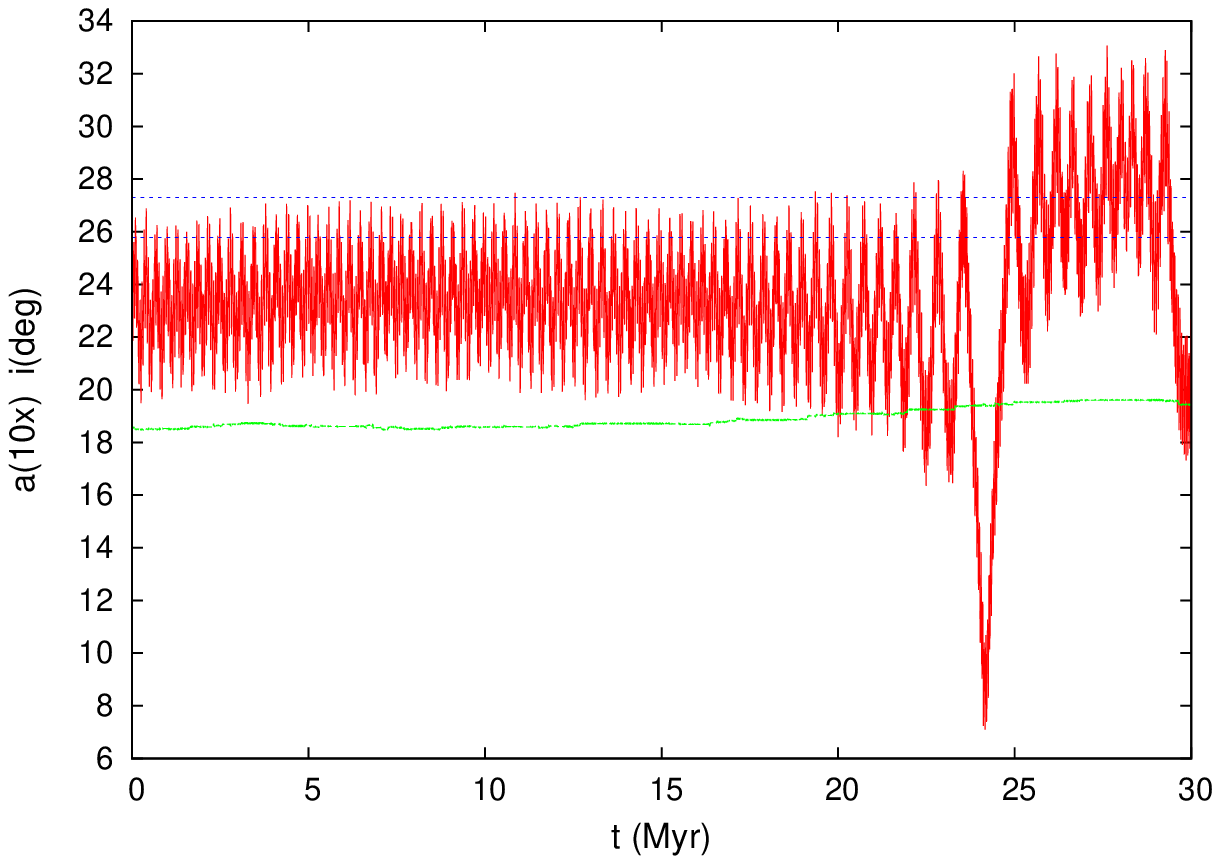}}
{\includegraphics[width=7.0cm,angle=0]{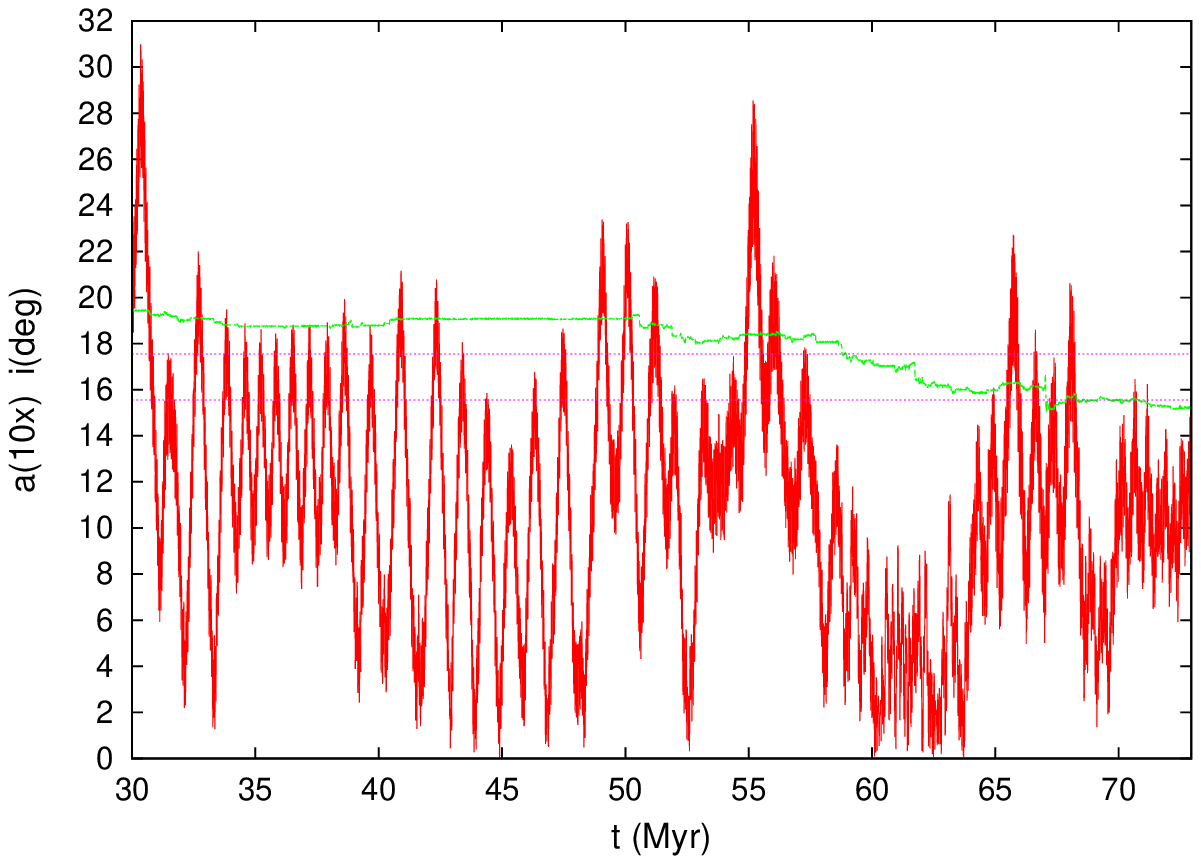}}
\caption{Dynamical evolution of a clone of the asteroid 2002 RN$_{137}$ later captured by 
Mars. The 2 horizontal lines of the upper panel represent approximately the region of influence of the 
secular resonance $g_5$ for that value of semi-major axis between
 25~Myr and 30~Myr.
 The 2  horizontal lines represent approximately the region of influence of the 
secular resonances $g_3$ and $g_4$,
 for that value of semi-major axis between  34~Myr and 38~Myr. On the y-axis 
inclination in degree and semi-major axis in astronomical units times 10.}
\label{secmil}
\end{figure}

\begin{figure}
\centerline
{\includegraphics[width=7.0cm,angle=0]{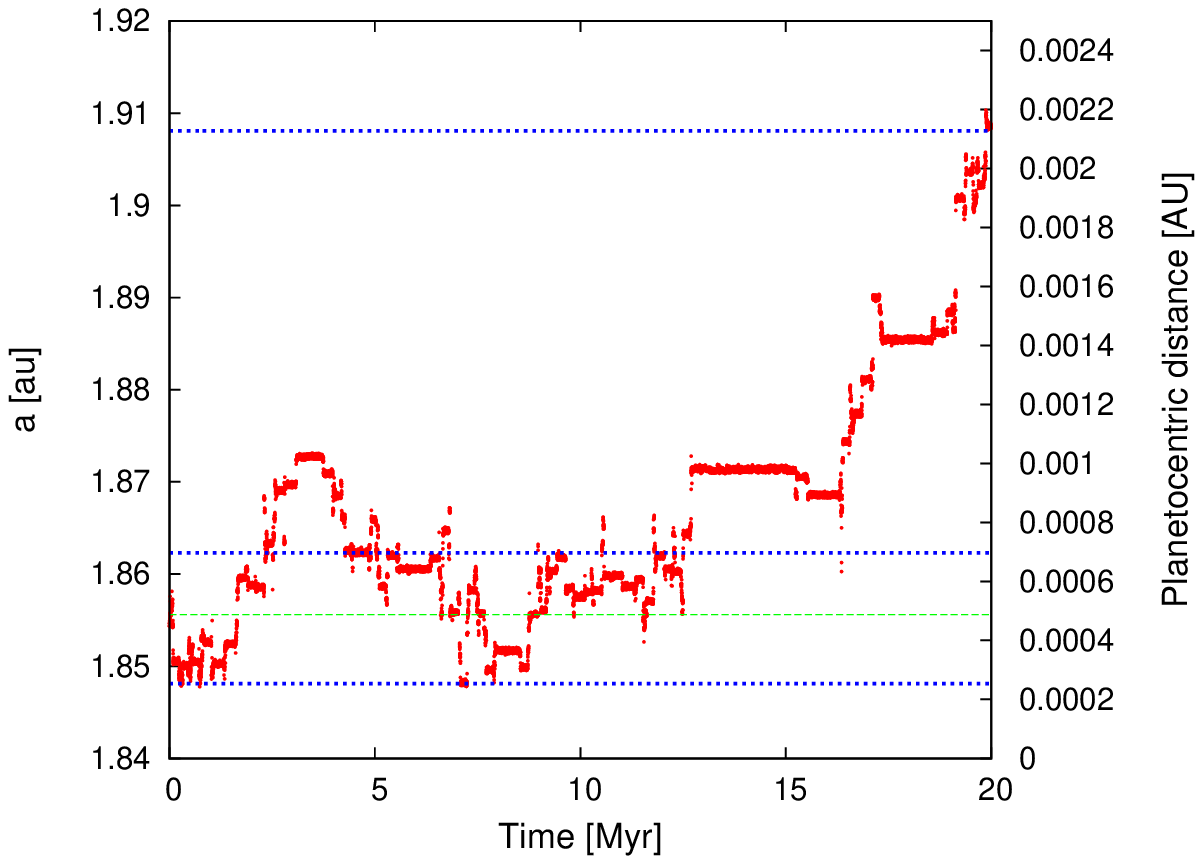}}
{\includegraphics[width=7.0cm,angle=0]{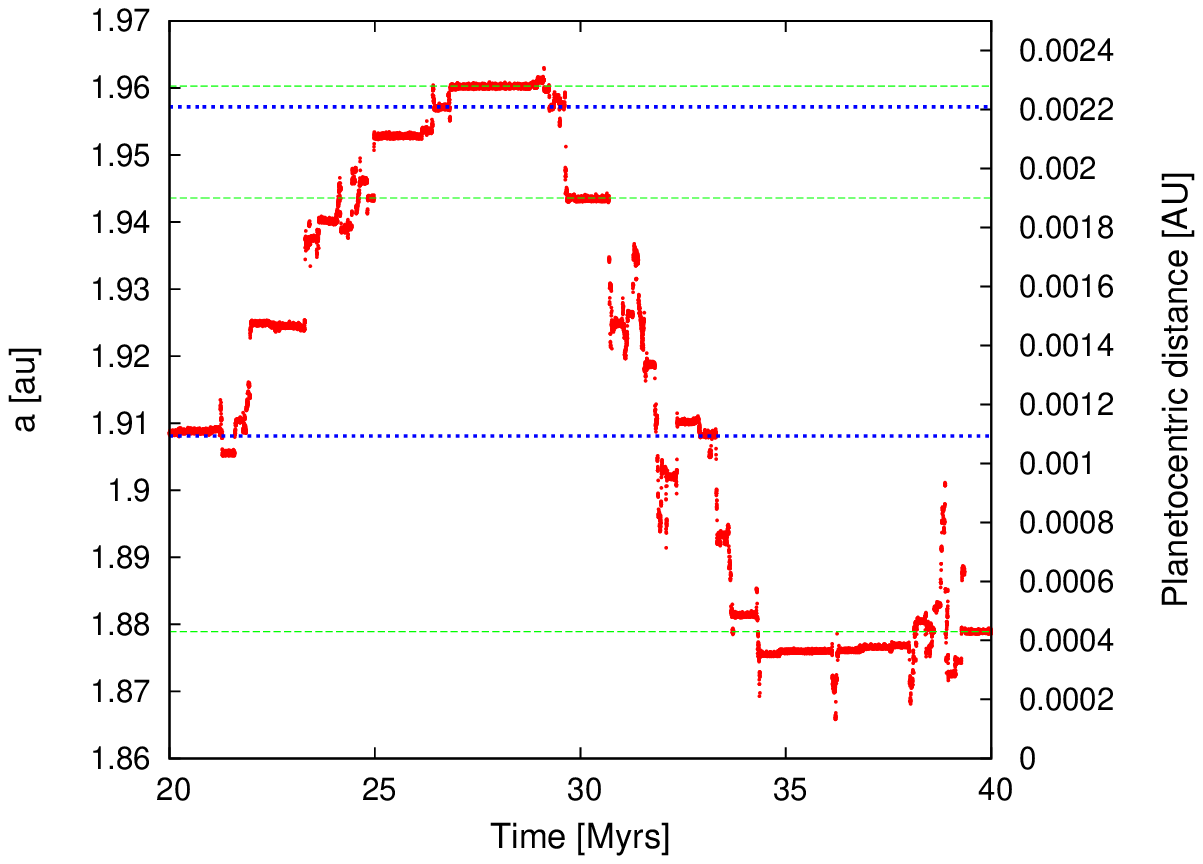}}
{\includegraphics[width=7.0cm,angle=0]{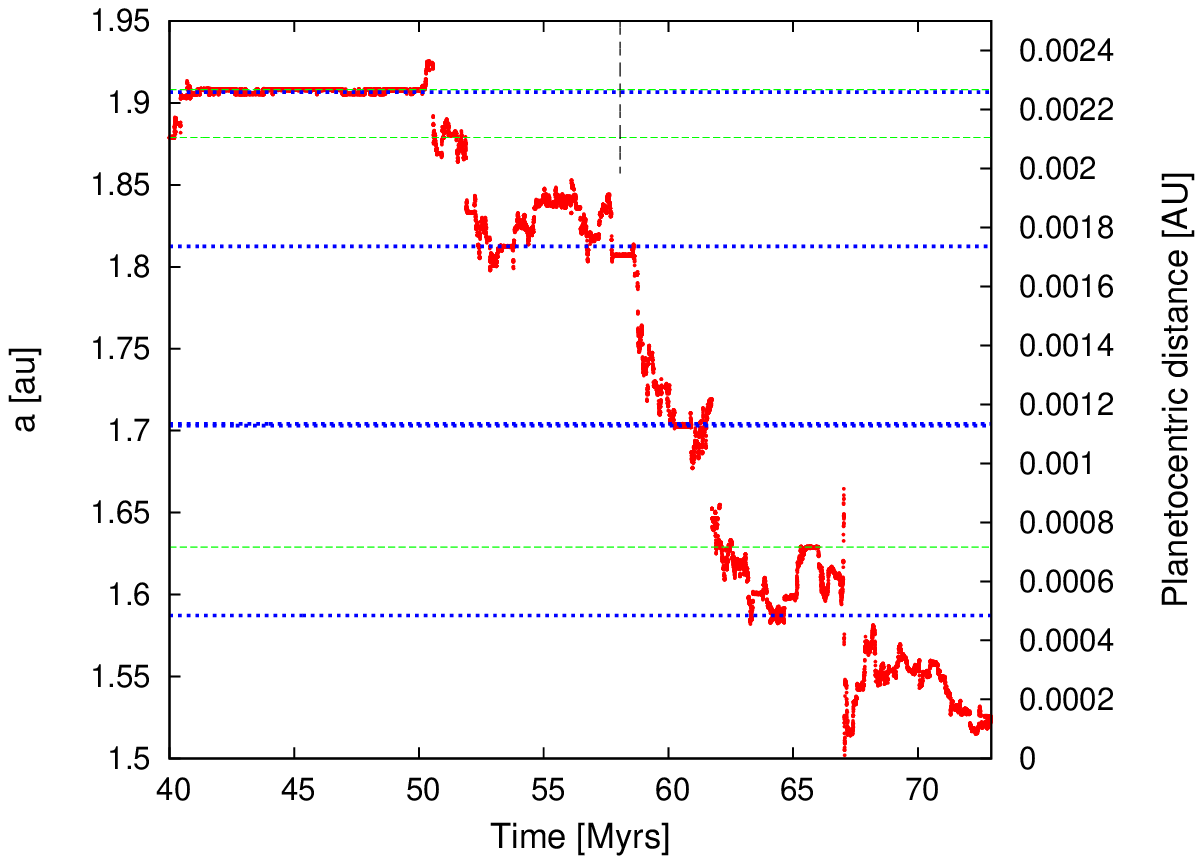}}
\caption{ Resonances and close encounters: dynamical evolution of a clone of the asteroid 2002 RN$_{137}$ later captured by 
Mars, divided in 3 mains parts, from 0~Myr to 20~Myr, from 20~Myr to 40~Myr and then till 60~Myr, when 
the changes in the orbit are dominated only by close encounters and resonances have secondary importance.
Position of the resonances: E2:1 = 1.5872 (it means a 2:1 resonance with the Earth is at 1.5872 au), J17-S14-2 = 1.6290 (it means a 
17-14-2 3bodyMMR with Jupiter, Saturn and the asteroids centered at 1.6290 au), M11:13 = 1.7031,J16:13 = 1.7042, M3:4 = 1.8458,
S12:1 = 1.8481, J13-S-9-2 = 1.8556, J20-S15-3 = 1.8623, J5-S1-1 = 1.8789, M5:7 = 1.9067, J13-S10-2 = 1.9081, J6-S4-1 = 1.9436, 
J13:3 = 1.9572, J19-S15-3 = 1.9603.}
\label{mmrs2}
\end{figure}

There are different possibilities how the clones get captured
 into co-orbital motion,
 an example of a capture into co-orbital motion is the candidate 2002 RN$_{137}$.
The description of the orbital evolution of one of its clones can help us to
understand the co-orbital evolution of the HCOs. A clone of 2002 RN$_{137}$
becomes a satellite of Mars after 73.237 Myr of integration
 and it stays like this for 6.5 kyr.
 We check its orbital evolution: 

\begin{itemize}
\item The close encounters, which change significantly the orbit of the 
asteroid and consequently its osculating elements, but in particular the 
semi-major axis. As shown in Tab.~\ref{tabREAL}
 the Hungaria fugitives have larger
 inclinations (in comparison with the initial
conditions of the asteroid families in the main belt),
which will lead to an escape from that region, because of SRs,
 and later on to close encounters with the 
terrestrial planets (e.g. Mars or the Earth).
In general this fact increases
 the possibility that the asteroid get captured
 into co-orbital motion. This
 was also shown for different initial conditions by \cite{Sch2012}.
The Hungaria candidate 2002 RN$_{137}$ represents the orbital behavior
 which we described previously.  
This decrease of the inclination favors the capture into co-orbital motion
 with Earth. 
Fig.~\ref{2002aei} let us see 
 multiple close approaches to Mars and after that also to the Earth.
 In the time-span between about 55 Myr and 65 Myr
 of integration, many 
close encounters are found, thus
 the inclination change dramatically and that leads to the
 Earth asteroids capture:
 the orbital elements of the captured asteroid lies
 in the stability window for that planet as shown by~\cite{Tab2000}.  
\item The resonances: MMRs, 3BMMRS and SRs, which change the eccentricity and
 again the inclination. An example of the most important resonances
 for this case are visible in the evolution of one of our fugitives:
  from about 25 Myr to 30 Myr,
 $g_5$ is active. The asteroid
 is inside the region of influence of this secular resonance
 (upper panel of Fig.~\ref{secmil}),
 having the inclination between $i\sim24^\circ$ and $i\sim32^\circ$
and keeping its
 semi-major axis between 1.9 au and 2.0 au. Then the asteroid travels
into the regions of influence of the SRs $g_3$ and $g_4$
 ~\cite[see also][, where these regions are
 well described]{War2009,Mil2010} from 
about 34 Myr and 38 Myr, where it does not have close encounters
 (Fig.~\ref{2002aei}, upper panel and ~\ref{secmil},  bottom panel).\\
The strongest MMRs and 3BMMrs which influence the orbit of
 this Hungaria appear to be:
 initially S12:1 and
 J13-S9-2 (where J is for Jupiter, S for Saturn and the last number
 is for the asteroid), J20-S15-3 and J13-S10-2 (Fig.~\ref{mmrs2}).
 Then from 20 Myr to about 22 Myr, the first order 3BMMR J13-S10-2 is active 
on the asteroid.  More over J19-S15-3 acts together with the
 $g_5$ for about 5 Myr from 25~Myr to 29.5 Myr and,
 from 29.5 Myr to about 30.5 Myr, 
we have J6-S4-1. From about 40 Myr to 50 Myr, M5:7,
 then from 51 Myr to 52 Myr,
 J5-S1-1 and in the end between 59 Myr and 66 Myr, when there are no close 
encounters, in chronological order M11:13 and J16:13 (for 2~Myr),
 E2:1 (for 1~Myr)
 and J17-S14-2 (for about 1.5~Myr).
All these resonances change significantly the eccentricity and the inclination
 facilitating close encounters of the asteroid with the planets and thus
 contributing to the change of osculating elements,
 in favor of some possible co-orbital orbit. 
\end{itemize}

\subsection{Sources of co-orbital bodies (results)}
\label{result}
\subsubsection{Population distributions}

We find that 3.3 \% of all the clones
 of all the fugitives (11) become HCOs: 
 1.8\% for Mars and 1.5\% the Earth, see Tab.~\ref{cratW3} for
 the distribution of the different classes and Tab.~\ref{probco} for the
 probability of becoming an HCO for each single Hungaria fugitive. 
This percentage\footnote{The percentage is the total number of co-orbital
 bodies per planet divided
the total number of asteroids (clones) integrated per region times 100, in this
 way we can compare better the 2 results.} represents the capture probability
 which we calculate from the total 
number of clones (a summary for the different co-orbital classes
 are given in Tab~\ref{cratW3}).
We obtain more Hungaria co-orbital bodies for Mars compared to
 those of the Earth,
 even if the difference is not so significant
 and many escapers experience different
 types of co-orbital motions. The QS class turns out to be
  the most favorable type of co-orbital motion.

{\bf We find more Hungaria Trojans in $L_5$ than $L_4$} and this is
what which was presently observed for real Trojans. 
We can conclude that 0.6\%
 Hungaria fugitives get captured in $L_4$ and 1.1\% in
 $L_5$ for Mars; for the Earth, 0.4\% Hungaria fugitives
 get captured in $L_4$ and 0.6\% in $L_5$.
 
Also a few cases of Hungaria Jumping-Trojans  are found
 and usually the Hungaria Jumping-Trojans stay in this condition for
 longer times. The maximum life is for a clone of 2002 RN$_{137}$,
 whose life-time is 58 kyr (see also Fig~\ref{lifinc}).

Some fugitives have the probability to become an HCO only for a single planet, 
i.e. 2001 XB$_{48}$ and 1996 VG$_9$ for the Earth or 1997 UL$_{20}$
 and 1999 UF$_5$ for Mars (Tab.~\ref{probco}).

The Hungarias with the highest probability to become a co-orbital asteroids
 have a probability of 8\% to be so and they are 
2002 RN$_{137}$ and 2000 SV$_2$. 2002 RN$_{137}$ 
is more likely to become a Mars HCO,
 the second one has both possibilities in equal measure
 (Mars or Earth HCO, Tab.~\ref{probco}).

\begin{table}
\begin{center}
\caption{Percentage of Hungaria captures in different classes (P. H.)
 from the total, average life time of the capture ($\bar{t}_{l}$).
 The * means that the
 total number of satellites is not equal to the sum of the total number
 of satellites and quasi satellites, because some clones can become QSs
 or satellites too, during their evolution.}
\begin{tabular}[h]{l|cc|cc|cc|cc|}
\hline
\hline
  Class      & P. H.   &       &  $\bar{t}_{l}$ [ky] &          \\
             &  Earth  & Mars  & Earth       & Mars   \\
\hline
\hline
Satellites      & 0.6 & 0.6 &  7.5$\pm$8.5  & 7.9$\pm$7.1\\
Quasi-satellites & 1.3 & 1.3 &  15.0$\pm$9.7 & 7.0$\pm$4.9 \\
Satellites (Subtot*)  & 1.9   & 1.9  &9.6$\pm$1.6 & 9.0$\pm$4.6   \\
\hline
\hline
Trojans & 1.1  &1.1& 7.9$\pm$3.9       &   10.7$\pm$4.4 \\
\hline
\hline
Horseshoes     &    0.0   &  0.4    & -  & 2.6$\pm$1.5\\
\hline
\hline
\end{tabular}
\label{cratW3}
\end{center}
\end{table}

\begin{table}
\begin{center}
\caption{Relative probability of each fugitive to become HCO (Tot.)
 and for each planet in percentage.}
\begin{tabular}{lccc}
\hline
Asteroid & Mars    & Earth  & Tot.\\
\hline
 (211279) 2002 RN137  & 2 & 6 & 8 \\
(152648) 1997 UL20   & 0 & 6 & 6\\
(141096) 2001 XB48   & 4 & 0 & 4\\
(24883) 1996 VG9    & 4 & 0 & 4\\
(41577) 2000 SV2    & 4 & 4 & 8 \\
(175851) 1999 UF5    & 0 & 2 & 2 \\
(39561) 1992 QA & 4   & 2 & 6 \\
\hline
\hline
\end{tabular}
\label{probco}
\end{center}
\end{table}

Table ~\ref{cratW} shows the distribution of asteroid captures (subdivided by
inclinations and total number too) found in this work and
 we compare partly our result with the work of \cite{Sch2012},
 ``partly'' because the initial conditions are different.
 The work of \cite{Sch2012} considered the region of 
the NEAs that covers also a small part of the Hungaria region. 
They called it region C, which considered this range of semi-major axis:
 1.54 au $< a <$ 2.20 au, but only at certain inclinations and eccentricities
 ~\cite[see][]{Sch2012}.

\begin{table}
\begin{center}
\caption{Captured asteroids from different region in percentage to the total.
 Regions are described in the text.
 $C^-$ is for $i<17^\circ$ and $C^+$ for $i>17^\circ$.
The numbers in the table are rounded to one digit.
 C is the region mentioned in \citet{Sch2012} and H
 stands for HCOs at their initial conditions.}
\begin{tabular}[!h]{|l||c|c||c|c||c|c|}
\hline
\hline
PLANET     &   $C^-$  & $C^+$&   $C_{tot}$     &   $H^-$  & $H^+$  & $H_{tot}$     \\
\hline

\hline
Earth   &  0.0     &   0.0 & 0.0  & 0.0 & 1.5 &1.5 \\
\hline
Mars   &  0.1      & 0.0  & 0.1  & 0.0  & 1.9 & 1.9\\
\hline
\end{tabular}
\label{cratW}
\end{center}
\end{table}

However much less HCOs were found 
in the work of \cite{Sch2012}, compared to us.
 This is because only certain peculiar regions in orbital elements
 can drive asteroids in close approaches with terrestrial planets, 
and even more peculiar ones give rise to asteroids in  co-orbital motions.

\subsubsection{Life time and orbits of the HCOs}

\begin{figure}
  \centering
\includegraphics[width=9.0cm]{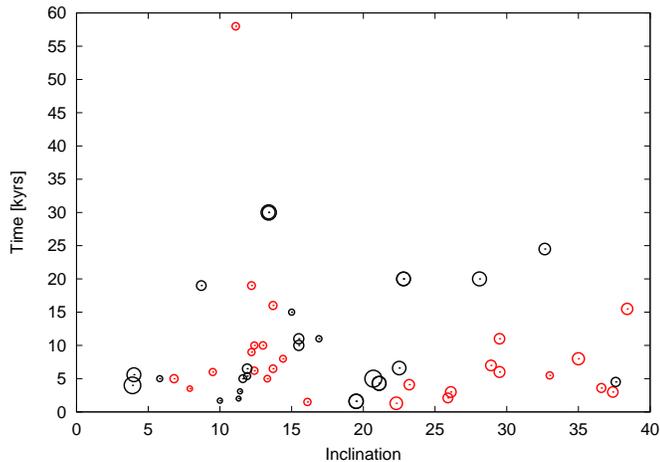}
\caption{Life time in thousand years versus inclination in degree. Earth HCOs
 are black and Mars HCOs are lighter. The size of the points depends
 among their eccentricity. The most stable asteroid is a Martian HCO with
 $a$=1.52368, $e$=0.3105 and $i$=11.11 ($a$, semi-major axis,
 $e$, eccentricity and $i$, inclination.}  
\label{lifinc}
\end{figure}

The HCOs have a short lifetime (or libration period),
 a mean of $\sim$ 10 kyr (9.6 kyr for the Earth
 and 9.0 kyr for Mars), with an exception for a jumping Trojan of
 2002 RN$_{137}$, which stays in this condition for more than  50 kyr.
 However HCOs have 
 lifetimes that usually range between 1 kyr and 20 kyr (Fig.~\ref{lifinc}).
  These results are in accordance with the life-time values found for 
real co-orbital asteroids, e.g. about 6.8 kyr for 2010 TK$_7$,
 a jumping-Trojan for the Earth, \citep{Con2011} or 1998 VF$_{31}$,
 an L$_5$ Mars Trojan, with a lifetime of 1.4 kyr \citep{deL2012}.
 These objects are usually transitional objects (with short dynamical
 life times) and the most stable HCOs have an inclination between $i=10^\circ$
and $i=17^\circ$, see also Fig.~\ref{lifinc}.\\
The Hungarias with smaller escape time from their original cloud have
a shorter lifetime as co-orbital objects. In fact the escape time
 (from the Hungaria region) decreases when
 they are perturbed by resonances, in particular SRs.
This is the case of asteroid 2002 SV$_2$, that is injected very soon into
 the terrestrial planets due to the $g_6$ \citep{Gal2013c}. 
Three clones of 2000 $SV_2$ become QSs (1 for the Earth and 2 for Mars),
e.g. one of them is captured very soon after 17.6 Myr of orbital
 evolution from initial conditions and it has its first close encounter
 with the Earth just at 10 Myr.

\begin{figure}
  \centering
\includegraphics[width=9.0cm]{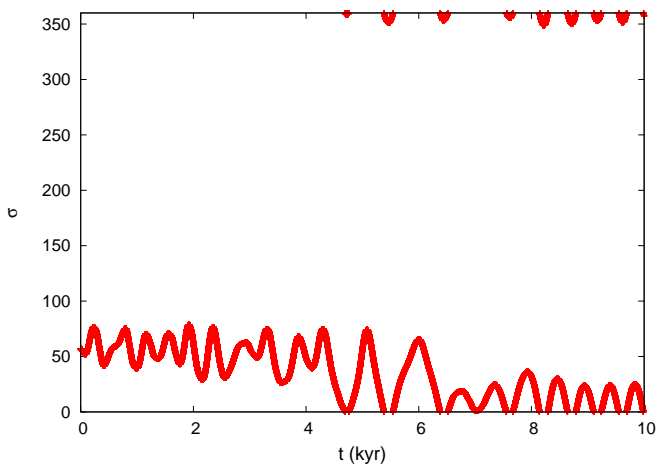}
\caption{Transition from $L_4$ to a QS-state, for the asteroid 2001 XB$_{48}$.
 The y-axis is the libration amplitude and the time is in thousand years.}  
\label{figtrans}
\end{figure}

\begin{figure}
  \centering
\includegraphics[width=9.0cm]{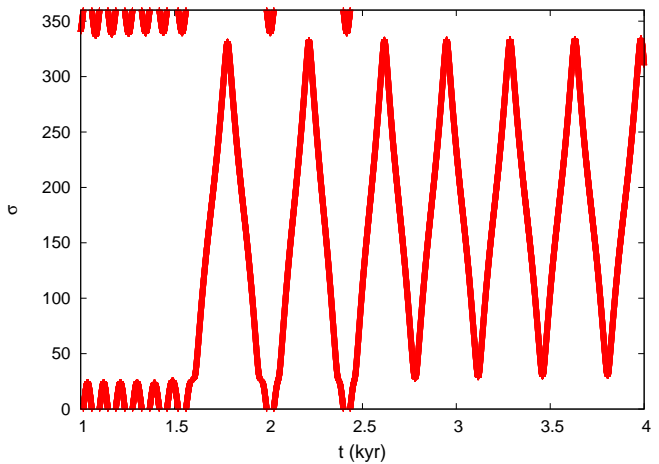}
\caption{Transition from satellite of the Earth to an
 Horseshoe orbit, for the asteroid 2002 SV$_{2}$.
 The y-axis is the critical angle and
 the time is in thousand years from the initial state as HCO.}  
\label{figtrans2}
\end{figure}

 Our investigation shows many asteroids can have
 multiple-captures, but never with different planets.
Many asteroids captured into HCOs can change their orbital behavior from QSs
 to tadpole orbits or into horseshoe orbits.
 The contrary is also possible, these multiple events was also
 found by~\cite{Wie1998,Con2002,Sch2012}.
 The switch of different
 types of co-orbital orbits happens especially for
 orbits with large eccentricity
 and/or high inclinations (and, as written before,
 Hungarias have these kind of orbits), having transitions from QS to
 horseshoe orbits~\citep{Nam1999},
 like 3753 Cruithne or 2002 AA29 \citep{Bra2004b},  co-orbital objects
 for the Earth. 
For instance, we detect
 for the Hungarias a clone of 2002 SV$_2$ and one of
 2001 XB$_{48}$  as transient co-orbital asteroids,
 see Fig.~\ref{figtrans} and ~\ref{figtrans2}.\\

From the dynamical point of view the variation in semi-major axis
 of the Earth HCOs (from now on EHCs) is larger than the one of Mars by about 
more than $\frac{1}{3}$ times, see Table~\ref{dispa}.\\
The inclination angles range from $\sim 3^\circ$ to very high inclined 
orbits, $\sim 40^\circ$.
The range for the inclination of the EHCs is in accordance
 with past studies,
 in fact stability windows for Earth-Trojans are covered and no cases
 are found between $24^\circ<i<28^\circ$. The inclination of the 
EHCs range on average between $15^\circ$ and $18^\circ$
 and for Trojan $\sim 16^\circ$ (Table~\ref{osculatio}).
Windows for Earth-Trojans, established by past works until now are:
 (a) $i<16^\circ$, (b) $16^\circ<i<24^\circ$ \citep{Tab2000}
 and (c) $28^\circ<i<40^\circ$ \citep{Dvo2012}.

Mars HCOs (from now on MHCs) have in general
 high inclined orbits close  to the original orbits, 
in fact they are less perturbed by close encounters compared to the EHCs.
 A remarkable thing is that the MHC  satellites have less inclined orbits 
than other type of MHCs, see Table~\ref{osculatio}.
The EHCs are usually less inclined than the MHCs, but the eccentricity is
 larger: $e_{EHC}\approx e_{MHC}/3$.
The eccentricity of Mars and Earth tad-pole orbits are similar, 
about $\sim0.32 < e < \sim0.36$ on the average.
 The Earth Hungaria MOs have larger eccentricities than
 tad-pole orbits of both planets
and also than MHC satellites (summarized in Table~\ref{osculatio}).  

\begin{table}
\begin{center}
\caption{Dispersion in semi-major axis during co-orbital motion.
 Sat. = satellites , QSs = Quasi Satellites, Troj. = Trojans,
 hors. = horseshoe orbits, $E$  = Earth and $M$ = Mars.
  $\Delta_{1,M}$ = maximum dispersion in 
semi-major axis for Martian HCOs and 
 $\Delta_{2,M}$ = minimum dispersion in semi-major axis for Martian HCOs.
 All measures are
 in units of $10^{-4}$ au.}

\begin{tabular}[h]{|l|c|cc|c|cc|}
\hline
\hline
  Class & $\bar{\Delta}_{E}$  & $\Delta_{1,E}$ & $\Delta_{2,E}$ & $\bar{\Delta}_{M}$   & $\Delta_{1,M}$  &  $\Delta_{2,M}$    \\
\hline
\hline
Sat.   & $17\pm18$  & 42 & 2       & $11\pm7$    & 19  & 5\\
QSs   & $26\pm3$  & 34 & 16       & $16\pm3$    & 22  & 11\\
\hline
Troj.   & $21\pm7$    & 29  & 7     &   $13\pm3$ & 18 & 6 \\
\hline
hors.  &   $18\pm2$   & 19  & 17   & -   & - & - \\
\hline
\hline
\end{tabular}
\label{dispa}
\end{center}
\end{table}

\begin{table}
\begin{center}
\caption{Orbital ranges for HCOs. EHCs = Earth Hungaria Co-orbital
 objects and MHCs = Mars Hungaria Co-orbital objects.
 $\bar{T}_p$ and $\bar{T}_j$ are respectively
 the Tisserand parameter relative to the planet in case (Mars and the Earth)
 and to Jupiter.
  $\bar{a}_{E}$ = average semi-major axis for EHCs, 
$\bar{e}_{E}$ = average eccentricity for EHCs and
  $\bar{i}_{E}$ = average inclination for EHCs. Sat. = satellite,
 QSs = Quasi Satellites and Troj. = Trojans.}
\begin{tabular}[h]{|l|c|c|c|}
\hline
\hline
  Class      &  $\bar{a}_{E}$ [au] & $\bar{e}_{E}$    &  $\bar{i}_{E}$  \\
             &  $\bar{T}_j$   &  $\bar{T}_p$   &    \\
\hline
\hline
Sat. & $0.9999\pm0.0011$& $0.431\pm0.010$  & $17.5\pm0.9$\\
       &  $5.942\pm0.007$  &   $2.687\pm0.007$  &          \\
QSs  & $1.0000\pm0.0013$& $0.546\pm0.012$  & $15.4\pm1.0$\\
       &  $5.912\pm0.007$  &   $2.618\pm0.004$  &          \\
\hline
Troj.   & $1.0000\pm0.0011$    & $0.321\pm0.004$  & $15.8\pm0.2$ \\
       & $6.010\pm0.005$   &  $2.842\pm0.001$   &          \\
\hline
EHCs   & $1.0000\pm0.0013$ & $0.507\pm0.011$  & $16.1\pm1.0$ \\
       &  $5.923\pm0.007$ & $2.643\pm0.005$ &          \\
\hline
\hline
  Class      &   $\bar{a}_{M}$ [au] & $\bar{e}_{M}$    &  $\bar{i}_{M}$  \\
             &  $\bar{T}_j$   &  $\bar{T}_p$   &    \\
\hline
\hline
Sat.   & $1.5236\pm0.0006$ & $0.331\pm0.006$   & $14.9\pm0.7$ \\
       & $4.411\pm0.002$  & $2.842\pm0.002$   &           \\
QSs   & $1.5238\pm0.0014$ & $0.387\pm0.015$ & $20.3\pm0.8$ \\
       &  $4.374\pm0.002$  & $2.774\pm0.005$  &           \\
\hline
Troj.   & $1.5238\pm0.0005$ & $0.355\pm0.010$  & $20.8\pm0.8$ \\
       &  $4.381\pm0.004$ & $2.792\pm0.006$  &          \\
\hline
MHCs   & $1.5238\pm0.0012$ & $0.373\pm0.012$ & $22.1\pm0.8$   \\
       & $4.362\pm0.002$ & $2.751\pm0.006$  &          \\
\hline
\end{tabular}
\label{osculatio}
\end{center}
\end{table}

The Tisserand parameter\footnote{$T_{body}= (1/2a)+\sqrt{a(1 - e^2)}cos(i)$,
 where a is the semi-major axis of the asteroid orbit, e is the eccentricity and i is the inclination} (Table~\ref{osculatio}) 
shows different values for each kind of co-orbital objects, in general the $T_j$ (J as Jupiter) of the QSs is usually higher. 
The typical $T_j$ of the EHCs ranges around $T_J=5.923$  and it is higher 
than the one of the MHCs, $T_j=4.362$.
Then the Tisserand parameter compared to the relative planet (but also in 
 respect to Jupiter) is higher for the MOs than the QSs.\\

\begin{figure}
  \centering
\includegraphics[width=7.0cm]{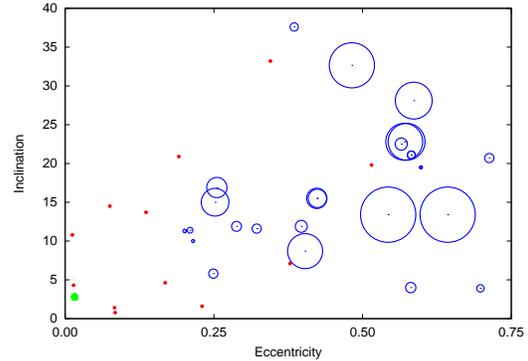}
\caption{Average (during its co-orbital motion) orbital elements of the
 Earth HCOs in comparison with
the real Earth co-orbital asteroids (circles with a dot inside):
 eccentricity versus inclination. The diameter of the circle is correspondent
to their life time (only for the HCOs). Earth is represented by the
 largest dot with the least eccentricity and the least inclination.}  
\label{cooE}
\end{figure}

\begin{figure}
  \centering
\includegraphics[width=7.0cm]{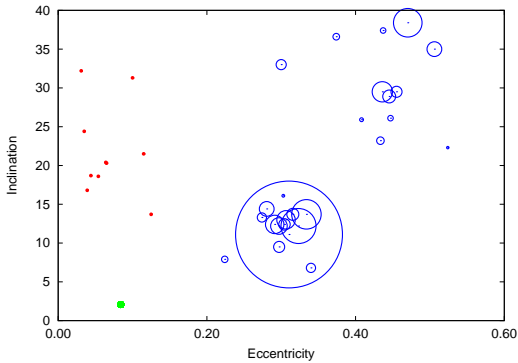}
\caption{Average  (during its co-orbital motion) orbital elements of the
 Mars HCOs in comparison with
the real Mars co-orbital asteroids (circles with a dot inside):
 eccentricity versus inclination. The diameter of the circle is correspondent
to their life time (only for the HCOs). Mars is represented by the
 largest dot with the least inclination.}  
\label{cooM}
\end{figure}

Some real co-orbital objects of the terrestrial planets could be of Hungaria
 origin as shown in Tab.~\ref{tabREAL} and ~\ref{osculatio}.
We compare the osculating elements (the average ones during
 the libration life, Fig.~\ref{cooE} and \ref{cooM}) of the HCOs with
 the real co-orbital asteroids described in Table ~\ref{tabREAL}.
  The HCOs favor co-orbital bodies 
with large eccentricity and indicate that it could be possible to find 
co-orbital bodies for Mars and Earth at large inclinations too, even more
 than $i=30^\circ$. This result -- displayed in Fig.~\ref{cooE} and~\ref{cooM} --
 seems to assert that co-orbital bodies
 which have a range in inclinations of $5^\circ<i<40^\circ$ and
 large eccentricities $0.22<e<0.53$ (even if most of the Mars HCOs
  are between 0.2 and 0.4) can be of Hungaria origin.
The majority of the HCOs lie between $i=5^\circ$ and $i=17^\circ$.
In this case the most probable former Hungarias are:
 Cruithne, 2006 FV$_{35}$, (85770) 1998 UP1 and YORP for the Earth and,
 due to their inclination,  1999 UJ$_7$, 1998 VF$_{31}$
 and 2011 SL$_{25}$ for Mars. 

Concerning the physical characteristics of the known co-orbital bodies 
for terrestrial planets,  some spectral type of these are known:
 for the Earth, Cruithne is a Q/S type\footnote{From The Near-Earth Asteroids
 Data Base (TNEADB) at earn.dlr.de/nea/table1\_new.html)},
 YORP is S/V type (from TNEADB); 
for Mars, 1997 $UJ_7$ is an X type, 1998 VF$_{31}$ is an S type:
 \cite{Riv2003} says it is an S(I) type (angrite)
 and \cite{Riv2007} suggest also for a S(VII) type (achondrite similar
 to the spectrum of 40 Harmonia). 
So Cruithne and 1998 VF$_{31}$ can
 again still be considered as possible HCOs,
 because Hungarias have some S-type asteroids, not the
 majority, but still 17\% 
~\citep{War2009} and especially 1997 UJ$_7$ which is an X-type 
asteroid like the majority of the 
 Hungarias \citep{Car2001,War2009}, even if not specified for the sub-group
 Xe-type and further spectroscopic analysis would be needed.

Considering the size of
the terrestrial planets' co-orbital bodies,
 we know the sizes which range from 
 the very small 2013 BS$_{45}$ of about 10-40 m to Cruithne of about 3.3 km
 in approximate diameter,
 but usually they are less than 1 km, similarly to the standard range
 of the Hungarias.

\newpage

\section{Conclusions}
\label{conclusions}
The capture of Hungarias as Earth or Martian Trojans is not only due
 to the migration of planets, but also to migration of asteroids 
from the Main Belt, this is proved
 physically by \cite{Riv2003},
 who show that two Martian Trojans are collision fragments of a larger body.
 Numerically, this migration towards the terrestrial
 planet was described by \cite{Gal2013a}, which emphasize the
 gravitational perturbations. Then, even \cite{Cuk2014} 
 show in particular the delivery of the aubrite
 meteorites by the Hungaria family, using also non-gravitational forces
 in the computation of the possible orbits.

The existence of HCOs have a low probability, 3.3\% of all Hungaria fugitives,
 but nevertheless the contribution of the Hungaria region is
 important in order to give rise to co-orbital objects for terrestrial
 planets and so the Hungaria region is one of the source-regions of
 the Main Belt for this kind of bodies.
  The capture of possible co-orbital Mars objects,
 is about 1.8\%, and for the Earth, it is 1.5\% of the total amount of
 the clones of the Hungaria fugitives in 100 Myr of evolution.
The Hungarias which have the highest probability (8\%) to become
 co-orbital objects of terrestrial planets are 2002 RN$_{137}$
 and 2000 SV$_2$.
The first time they become co-orbital objects on average is
 at $\sim70$ Myr after their orbital evolution from the original
 (present) position.
 The HCOs majority become QSs and concerning the former Hungarias
 captured into tad pole motions, they will be captured around $L_5$.
 We found less captures for $L_4$ Trojans 
for both planets and we did not find any Venus HCOs,
 in agreement with the present observations.
 There are some cases of Jumping Trojans and
 with the longest life time, the maximum detected life time
 is 58 kyr.
 Also many HCOs behave like transitional co-orbital objects.
Some Hungarias can become co-orbital objects of both planet together,
i.e. 2000 SV$_2$ and some exclusively of only one
 like 1997 UL$_{20}$ and 1999 UF$_5$ of Mars and,
 2001 XB$_{48}$ and 1996 VG$_9$ of the Earth.

The mechanism found in this work to transport the asteroids
 from the Hungaria region close to the terrestrial
planets and finally captured into co-orbital motion are the following ones:
\begin{enumerate}
\item close encounters with Mars and Earth. Especially for the Earth case,
 the close encounters decrease the inclination
 in such a way that the Hungarias enter the window of stability
 for the Earth co-orbital objects as shown in \cite{Tab2000}.
\item resonances: SRs, such as $g_5$ and $g_6$; MMRs such as M3:4, M11:13
 and J16:13 and 3BMMRs, such as J13-S10-2 and J5-S1-1.
\end{enumerate}

The average libration period of the HCOs is quite short
$\sim10 kyr$ (9.6 kyr for the Earth and 9.0 kyr for Mars),
 in accordance with the real co-orbital objects
 of terrestrial planets, \citep[e.g. 6.8 kyr for 2010 TK$_7$][]{Con2011}.
Furthermore our investigations show that the
 Hungarias with the shortest lifetime
 are the first ones to escape from the Hungaria cloud,
 i.e. 2000 $SV_2$, Tab.~\ref{HCOs}.

Mars is the planet which captures more Hungarias,
 because of the shorter distance, even if the difference
 between Mars and the Earth capture probability is relatively small.
 Probably this can explain the smaller number of Earth
 co-orbital asteroids compared to Mars,
because many asteroids will be captured by Mars
 (close encounters make the bodies achieving
 too large eccentricities to become
 Earth co-orbital bodies).
 However, the evolution of other families of the main belt
 in this sense should be studied in more detail in the future.

Concerning the HCOs' orbits,
they range from $i\sim3^\circ$ to $i\sim40^\circ$. 
The EHCs average inclination range is $15^\circ<i<18^\circ$
 and for Trojans $i\sim16^\circ$; this is in agreement with \cite{Sch2012}. 
The high inclined HCOs are favourable for MHCs instead
 of for EHCs and among the MHCs the satellites have the lowest
 inclined orbits ($14.9^\circ$).
The eccentricity of the EHCs is
on average 3 times the one
 of the MHCs, only for the tad pole orbits, it is similar.
The EHCs satellites have $e=0.52$, and the Martians $e=0.37$.

The typical Tisserand parameter with respect to Jupiters for
 EHCs is $T_j=5.923$ and for MHCs is less, $T_j=4.392$ and for satellites
 $T_j$ is higher than the QSs. 

Some real co-orbital asteroids have orbits which have a high probability 
 to be former Hungarias, like Cruithne, (277810) 2006 FV$_{35}$, 
 (85770) 1998 UP1 and YORP for EHCs and
 (101429) 1998 VF$_{31}$ and, (88719) 2011 SL$_{25}$ for MHCs. 
 In particular Cruithne and 1998 VF$_{31}$ have
 both orbital elements and physical characteristics typical of
 the S-type HCOs.
 Further investigations have to be done to look for
 the origin of present co-orbital objects of the 
 terrestrial planets, both dynamical and observational studies.
 In the next work we will perform a new study
 for HCOs taking into account also non-gravitational forces.

\section*{Acknowledgments}
R. Schwarz and M. A. Galiazzo want
to acknowledge the support by the Austrian 
FWF project P23810-N16 and thank
 Barbara Funk, \'Akos Basz\'o and Simone Recchi 
for some suggestions on the text.

\label{lastpage}
\end{document}